\documentclass[intlimits,twoside,a4paper]{article}

\usepackage{amsmath,amssymb}
\usepackage{graphicx}
\usepackage{subfig}
\usepackage[T2A]{fontenc}
\usepackage[cp1251]{inputenc}

\usepackage[eqsecnum]{cmpj3}

\issue{2020}{23}{3}{33705}
\doinumber{10.5488/CMP.23.33705}
\title[Ab initio study of structural, electronic and magnetic properties]%
{Ab initio study of structural, electronic and magnetic properties of XSn$_3$ (X = Gd, Cm) and Gd$_x$Cm$_{1-x}$Sn$_3$ compounds%

}

\author[M. Adnane \textsl{et al.}]{M. Adnane\refaddr{label1}, L. Djoudi\refaddr{label1,label2},
       M. Merabet\refaddr{label1,label2}, M. Boucharef\refaddr{label1}, F. Dahmane\refaddr{label2},  S. Benalia\refaddr{label1,label2}, M. Mokhtari \refaddr{label2,label3}, D.  Rached\refaddr{label1}}
\addresses{
\addr{label1} Laboratoire des Mat\'eriaux Magn\'etiques, Universit\'e  Djillali Liab\'es de Sidi Bel-Abb\'es, Alg\'erie.
\addr{label2} D\'epartement de SM, Institut des Sciences et des Technologies, Centre Universitaire de Tissemsilt, 38000 Tissemsilt, Algeria.
\addr{label3} Universit\'e des Sciences et de la Technologie d'Oran Mohamed Boudiaf, USTO-MB, LEPM, BP 1505, El M' Naouar, 31000 Oran, Algeria.  }

\date{Received March  26, 2019, in final form April 26, 2020}

\begin{document}

\maketitle
\begin{abstract}
In this paper, the structural, electronic and magnetic properties of the GdSn$_3$, CmSn$_3$ and Gd$_x$Cm$_{1-x}$Sn$_3$ compounds ($x = 0.25$, $0.5$ and $0.75$) were studied using the full-potential linearized augmented plane wave method, within the generalized gradient approximation$+U$. The ground-state properties are determined for the bulk materials GdSn$_3$, CmSn$_3$ and Gd$_x$Cm$_{1-x}$Sn$_3$ crystallized in AuCu$_3$-type structure. The calculated structural, electronic and magnetic properties of GdSn$_3$ compound are in good agreement with the existing experimental and theoretical data. It is found that the most stable magnetic configurations of both compounds CmSn$_3$ and GdSn$_3$ are anti-ferromagnetic type A (AFM-A) and have a metallic behavior. The magnetic moment found decreases with increasing the Cm composition in Gd$_x$Cm$_{1-x}$Sn$_3$ compounds. The results show that the GdSn$_3$, CmSn$_3$ and Gd$_x$Cm$_{1-x}$Sn$_3$ compounds share some properties, and may well be useful for spintronic applications.

\end{abstract}

\section{Introduction}
In the last few years, rare earth intermetallic compounds and their alloys have attracted a great deal of attention  being among the most important material systems used in many fields, including optoelectronic, electronic and magnetic applications \cite{1, 2, 3, 4, 5, 6}.  These materials are usually found in several electronic products (cell phones, laptops, cameras, plasma screen TVs, solar panels...) because of their magnetic properties \cite{7}. These materials have many interesting technological applications such as high-density magnetic recording and magnetic refrigeration.
The intermetallic compounds based on rare earth and transition metals, show many interesting properties. Among their main properties, we find the presence of different magnetic structures, valence instabilities and magnetic moment formation \cite{8, 9}. The interactions between the ``f'' states, which are highly correlated and greatly localized in the rare earth elements, and the ``d'' states of transition metal atoms have a direct effect on the above-mentioned properties.  The lanthanide and the actinide series transition metals belong to the f-block, where the 4f and 5f states have a direct and prominent effect on the physical properties of both compounds, respectively \cite{10}. For lanthanide series transition metals, the 4f states are confined deeply in the core and are generally considered to be localized. Being the electronic wave functions in actinides, they are more extensive than in lanthanide series transition metals \cite{11}.

Knowledge of the basic properties of the gadolinium tin alloy system (GdSn$_3$) and the curium tin alloy system (CmSn$_3$) and their alloys like other materials are of great technological interest. These properties help us to produce new materials, which are used more in high-tech applications. Due to the importance of these materials, several scientific researches have been done; e.g., Abraham et al. \cite{12}   have used the FP-LAPW to study the structural and electronic properties of GdSn$_3$ and YbSn$_3$ intermetallic compounds.  Shafiq et al. \cite{13}   also used the FP-LAPW to study this rare-earth intermetallic compounds, SmIn$_3$, EuIn$_3$, GdIn$_3$, SmSn$_3$, EuSn$_3$ and GdSn$_3$. In 2015,  Benidris et al. \cite{14}   used the same method (FP-LAPW) based on both GGA and GGA$+U$ approaches to study the structural and magnetic properties of RESn$_3$ (RE: rare-earth elements).  Abraham et al. reported that the GdSn$_3$ compound is ductile and metallic in nature.  Shafiq et al. reported that the gadolinium tin compound is anisotropic and elastically stable. Among the results, found by Benidris et al., is the statement that the GdSn$_3$ compound is anti-ferromagnetic of type~A (AFM-A). 

	As far as we know, there is no scientific research available on the structural, electronic and magnetic properties of CmSn$_3$ and Gd$_x$Cm$_{1-x}$Sn$_3$ alloys. In order to make full use of the properties of the bulk GdSn$_3$ and CmSn$_3$ and their Gd$_x$Cm$_{1-x}$Sn$_3$ alloys, a theoretical study of the structural, electronic and magnetic properties is obligatory. For this purpose, we study the ground state properties of GdSn$_3$ and CmSn$_3$ bulk and of the Gd$_x$Cm$_{1-x}$Sn$_3$ compounds in AuCu$_3$-type structure at zero pressure, using the FP-LAPW method, within the GGA$+U$ approximation within the density functional theory (DFT). The WIEN2K code software is used in this study \cite{15}. The presentation of the paper is as follows: in section~2, we described the employed method and the details of the calculations; results and discussions of the structural, electronic and magnetic properties of GdSn$_3$, CmSn$_3$ and Gd$_x$Cm$_{1-x}$Sn$_3$ compounds are presented in section~3; finally, conclusions and remarks are made in section~4.

\section{Computational method}

\begin{figure}[!t]
\centerline{%
\includegraphics[width=9 cm]{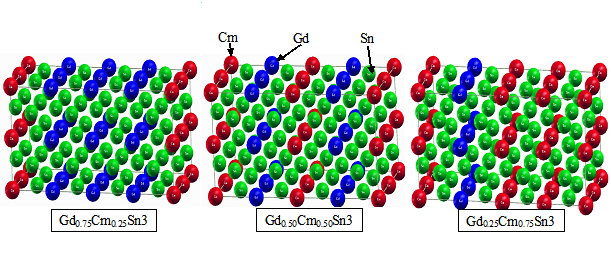}}
\caption{(Colour online) Geometry of Gd$_{0.75}$Cm$_{0.25}$Sn$_3$, Gd$_{0.50}$Cm$_{0.50}$Sn$_3$, and Gd$_{0.25}$Cm$_{0.75}$Sn$_3$ compounds.}
\label{Fig:F1H}
\end{figure}

In this study, the structural, electronic and magnetic properties of the  GdSn$_3$, CmSn$_3$ and Gd$_x$Cm$_{1-x}$Sn$_3$  compounds are investigated using the FP-LAPW simulation program approach based on the density functional theory; where we have used, as an approximation for the calculation of exchange-correlation energy functional, the well-known generalized gradient approximation (GGA)   \cite{16} and the GGA$+U$. This simulation is implemented in the WIEN2k code. In the FP-LAPW calculations, each unit cell is partitioned into non-overlapping muffin-tin spheres around the atomic sites. Basis functions are expanded in combinations of spherical harmonic functions inside the non-overlapping spheres. Core states are treated within a multi-configuration relativistic Dirac-Fock approach \cite{17}, while valence states are treated in a relativistic scalar approach \cite{18}. In the interstitial region, a plane wave basis is used and the expansion is limited with a cut-off parameter, $R_{\text{MT}}\cdot K_{\text{MAX}} = 9$ for all the compounds. $R_{\text{MT}}$ is the minimum radius (MF) of the sphere in the unit cell;  $K_{\text{MAX}}$ is the magnitude of the largest $K$ vector used in the plane wave expansion. The MT radii are adopted to be $2.85$, $2.96$ and $2.87$ Bohr for Gd, Sn, and Cm atoms. To model Gd$_x$Cm$_{1-x}$Sn$_3$, we take the 64-atom Gd$_n$Cm$_{8-n}$Sn$_{48}$ supercell, which corresponds to $2\times 2\times 2$ supercell. Geometry of Gd$_{0.75}$Cm$_{0.25}$Sn$_3$, Gd$_{0.5}$Cm$_{0.5}$Sn$_3$, and Gd$_{0.25}$Cm$_{0.75}$Sn$_3$ compounds are shown in fiqures~\ref{Fig:F1H}. However, it may be noted that the addition of Cm elements will necessitate the formation of a larger supercell, and hence calculations become intensive.

\section{Results and discussion}

\subsection{Magnetic phase stability and equilibrium structural properties}

\begin{figure}[!t]
\centerline{%
\includegraphics[width=7 cm]{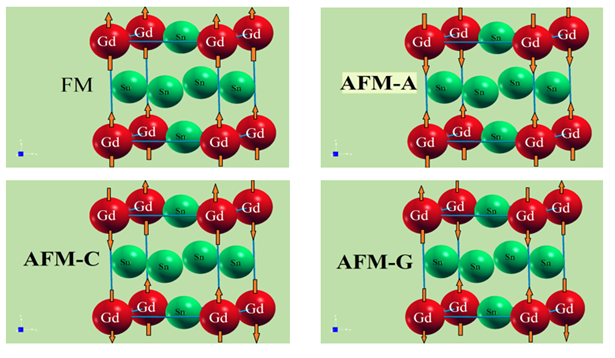}}
\caption{(Colour online) The possible magnetic configurations of GdSn$_3$ and CmSn$_3$ compounds.}
\label{Fig:F2H}
\end{figure}

\begin{figure}[!t]
\centerline{%
\includegraphics[width=7 cm]{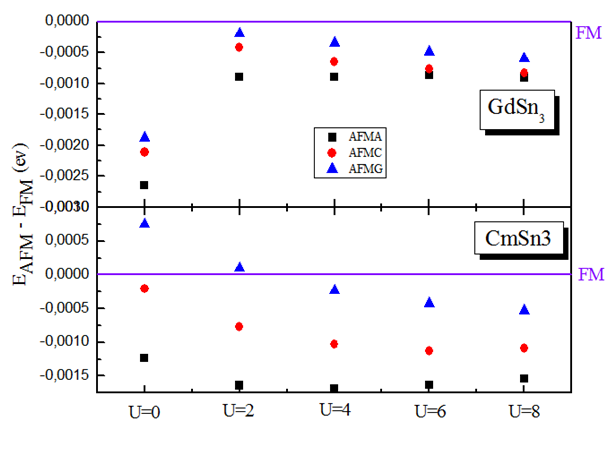}}
\caption{(Colour online) The magnetic stability as a function of the effective Coulomb interaction of GdSn$_3$ and CmSn$_3$ compounds using GGA$+U$ approach. The ferromagnetic state is taken as reference.}
\label{Fig:F3H}
\end{figure}

The ground-state structural parameters have been obtained by minimizing the total energy with respect to the volume, by fitting this total energy versus the volume data on the nonlinear Murnaghan equation of state \cite{19}. Firstly, we computed the magnetic phase stability of the GdSn$_3$ and CmSn$_3$ bulk and Gd$_x$Cm$_{1-x}$Sn$_3$ compounds ($x = 0.25$, $0.5$ and $0.75$) in their cubic structure, AuCu$_3$ structure, with the space group Pm3m (no. 221). We have several possible magnetic configurations that can be established, comprising four states: (a: FM) Ferromagnetic state, (b: AFM-A state) Anti-Ferromagnetic type A, (c: AFM-C state) Anti-Ferromagnetic type C and (d: AFM-G state) Anti-Ferromagnetic type~G, figure~\ref{Fig:F2H}. 
In order to estimate the magnetic phase stability of the GdSn$_3$ and CmSn$_3$ bulk, we  plotted in figure~\ref{Fig:F3H} the total energies as a function of the Hubbard potential using the GGA$+U$ approach, while the ferromagnetic state is taken as reference. It is very clear from figure~\ref{Fig:F3H} that the most stable magnetic configurations of both compounds GdSn$_3$ and CmSn$_3$ are anti-ferromagnetic type A (AFM-A). Therefore, we can predict the most stable magnetic configurations of these compounds: Gd$_{0.75}$Cm$_{0.25}$Sn$_3$,  Gd$_{0.50}$Cm$_{0.50}$Sn$_3$ and Gd$_{0.25}$Cm$_{0.75}$Sn$_3$, which are anti-ferromagnetic type A (AFM-A).  In table~\ref{tbl-smp1}, we  listed the magnetic stability for the  investigated of the systems and compared the results with the available experimental and theoretical data \cite{14, 20}. This result is in excellent agreement with the experimental and theoretical data for GdSn$_3$. To our knowledge, there is no other research for CmSn$_3$, Gd$_{0.75}$Cm$_{0.25}$Sn$_3$,  Gd$_{0.50}$Cm$_{0.50}$Sn$_3$ and Gd$_{0.25}$Cm$_{0.75}$Sn$_3$ compounds to make comparison, that is why our results remain purely predictive. We can assert that the local magnetic 4f moments of the GdSn$_3$ and 5f moments of the CmSn$_3$ are mainly responsible for their magnetism.
\begin{table}[!t]
\caption{ Magnetic phase stability as a function of the ective Coulomb interaction of GdSn$_3$, CmSn$_3$, Gd$_{0.75}$Cm$_{0.25}$Sn$_3$, Gd$_{0.5}$Cm$_{0.5}$Sn$_3$, and Gd$_{0.25}$Cm$_{0.75}$Sn$_3$ compounds compared with available experimental and/or theoretical data.}
 \label{tbl-smp1}
 \vspace{2ex}
 \begin{center}
 \renewcommand{\arraystretch}{0}
 \begin{tabular}{|c|c||c|c|c|c|c|c||}
 \hline
 GGA$+U$ &$U=0$&$U=2$&$U=4$&$U=6$&$U=8$&Exp.&Other\strut\\
 
 \hline
 \rule{0pt}{2pt}&&&&&\\
 \hline
 GdSn$_3$&AFM-A&AFM-A&AFM-A&AFM-A&AFM-A&AFM \cite{20}&AFM-A \cite{21} \strut\\
 
  \hline 
    
 CmSn$_3$&AFM-A&AFM-A&AFM-A&AFM-A&AFM-A&& \strut\\
 
  \hline 
 Gd$_{0.75}$Cm$_{0.25}$Sn$_3$&AFM-A&AFM-A&AFM-A&AFM-A&AFM-A&& \strut\\
 
  \hline 
 Gd$_{0.50}$Cm$_{0.50}$Sn$_3$&AFM-A&AFM-A&AFM-A&AFM-A&AFM-A&& \strut\\
 
  \hline 
 Gd$_{0.25}$Cm$_{0.75}$Sn$_3$&AFM-A&AFM-A&AFM-A&AFM-A&AFM-A&& \strut\\
 
  \hline 
 \end{tabular}
 \renewcommand{\arraystretch}{1}
 \end{center}
 \end{table}

In table~\ref{tbl-smp2}, we  summarized the calculated lattice constants, and the bulk module and its pressure derivative of GdSn$_3$ and CmSn$_3$ compounds together with the available experimental and theoretical data. As can be seen, the obtained lattice constant for GdSn$_3$ is in reasonable agreement with the theoretical data \cite{14}, and is $1.47 \%$  higher than the experimental value, which ensures the reliability of the present first-principles computations. For CmSn$_3$, we have no theoretical or experimental data to make comparison. The calculated lattice parameters of GdSn$_3$ and CmSn$_3$ compounds increase with an increase of the Hubbard potential. This increase  is due to the effect of the GGA approximation and the repellent effect of the Coulomb potential \cite{21}. We also notice  that the two calculated lattice parameters of GdSn$_3$ and CmSn$_3$  have almost the same values. It is very clear from table~\ref{tbl-smp2} that the calculated bulk modulus and its pressure derivative for both compounds varies non-linearly with an increasing Hubbard potential. 
 \begin{table}[!b]
\caption{Calculated structural properties: lattice parameter $a$ and bulk modulus $B$ and its pressure derivative $B'$ for of GdSn$_3$ and CmSn$_3$ compounds compared with available experimental and/or theoretical data.}
\label{tbl-smp2}
\vspace{2ex}
\begin{center}

\renewcommand{\arraystretch}{0}
\begin{tabular}{|c|c||c|c|c|c|c|c|}
\hline
 \multicolumn{3}{|c|}{GGA$+U$}  & $U = 0$  & $U = 2$ & $U = 4$  & $U = 6$  & $U = 8$ \strut\\
\hline
\rule{0pt}{2pt}&&&&&\\
GdSn$_3$& $a$ $\r{A}$&Pre. cal. & 4.74 & 4.74 & 4.74 & 4.75 & 4.75 \strut\\
\cline{3-8}
&&Other cal.  & 4.74 \cite{21}  & 4.74\cite{21}  & 4.74 \cite{21}  & 4.75 \cite{21}  & 4.75\cite{21}   \strut\\
&&  & 4.681 \cite{12}    &  &   & &   \strut\\
\cline{3-8}
&&exp. & 4.67 \cite{20} & -- & -- & -- & --  \strut\\
\cline{2-8}
&$B$(GPa)&Pre. cal. & 56.09 & 57.79 & 57.13& 57.19 & 57.10 \strut\\
\cline{3-8}
&&Other cal.  & 56.01 \cite{21}  & 56.58\cite{21}  & 56.98 \cite{21}  & 56.82 \cite{21}  & 5.72\cite{21}   \strut\\
&&& 58.03\cite{12}  & --  & --  &--  & --   \strut\\
\cline{3-8}
&&exp. & --& -- & -- & -- & --  \strut\\
\cline{2-8}
&$B'$&Pre. cal. & 4.460 & 4.326 & 4.571 & 4.411 & 4.491 \strut\\
\cline{3-8}
&&Other cal.  & 4.64 \cite{12}  & --  & --  &--  & --   \strut\\
\cline{3-8}
&&exp. & --& -- & -- & -- & --  \strut\\
\hline
CmSn$_3$&$a$ $\r{A}$&Pre. cal. & 4.73 & 4.75 & 4.76 & 4.77 & 4.78 \strut\\
\cline{3-8}
&&Other cal.  &   -- &  -- &  -- &  -- &  --  \strut\\
\cline{3-8}
&&exp. & --  & -- & -- & -- & --  \strut\\
\cline{2-8}
&$B$(GPa)&Pre. cal. & 55.83 & 56.06 &56.15 & 56.35 & 56.77 \strut\\
\cline{3-8}
&&Other cal.  &  --  &  -- &  -- &  -- &  --  \strut\\
\cline{3-8}
&&exp. & --  & -- & -- & -- & --  \strut\\
\cline{2-8}
&$B'$&Pre. cal. & 4.445 & 4.479 & 4.415 & 4.31 & 4.38 \strut\\
\cline{3-8}
&&Other cal.  &   -- & --  & --  & --  & --   \strut\\
\cline{3-8}
&&exp.&  -- & -- & -- & -- & --  \strut\\
\hline
\end{tabular}
\renewcommand{\arraystretch}{1}
\end{center}
\end{table}

\subsection{ Electronic and magnetic properties}

\subsubsection{ Band structure}
\begin{figure}[!b]
\centerline{%
\includegraphics[width=9 cm]{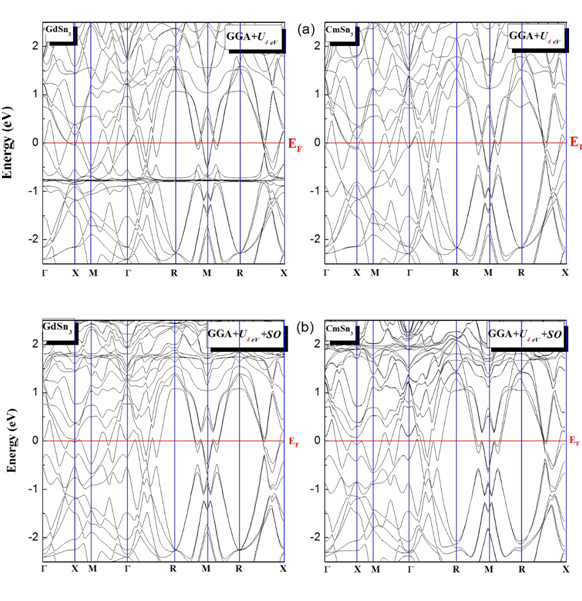}}
\caption{(Colour online) The calculated band structures of GdSn$_3$ and CmSn$_3$ using GGA$+U$ approach with and without spin-orbit coupling.}
\label{Fig:F4H}
\end{figure}

A good knowledge of electronic band structure in materials provides a valuable information concerning their potential utility in producing multi-electronic devices. In this context, we  calculated the electronic band structure properties of GdSn$_3$ and CmSn$_3$ in the cubic structure at these equilibrium lattice constants using the GGA$+U$ formalism. GGA$+U$ ($U = 4$~eV) approximation with and without spin orbit coupling (SOC) is used to treat the exchange-correlation effects in these systems in their magnetic phases. The calculated band structures of GdSn$_3$ and CmSn$_3$ compounds using GGA$+U$ ($U = 4$~eV) approach with and without the spin orbit (SO) are illustrated in figure~\ref{Fig:F4H} a and figure~\ref{Fig:F4H} b, respectively. It is apparent from figure~\ref{Fig:F4H} a  and figure~\ref{Fig:F4H} b that the GdSn$_3$ and CmSn$_3$ are of a metallic nature. From figure~\ref{Fig:F4H} b, it is very difficult to identify the effect of spin orbit, except for a small shift in energy levels in the energy band. Calculations yield similar band structures for GdSn$_3$ and CmSn$_3$, and this reflects their similar properties.

\subsubsection{Total and partial densities of states}

\begin{figure}[!b]
\centerline{%
\includegraphics[width=9 cm]{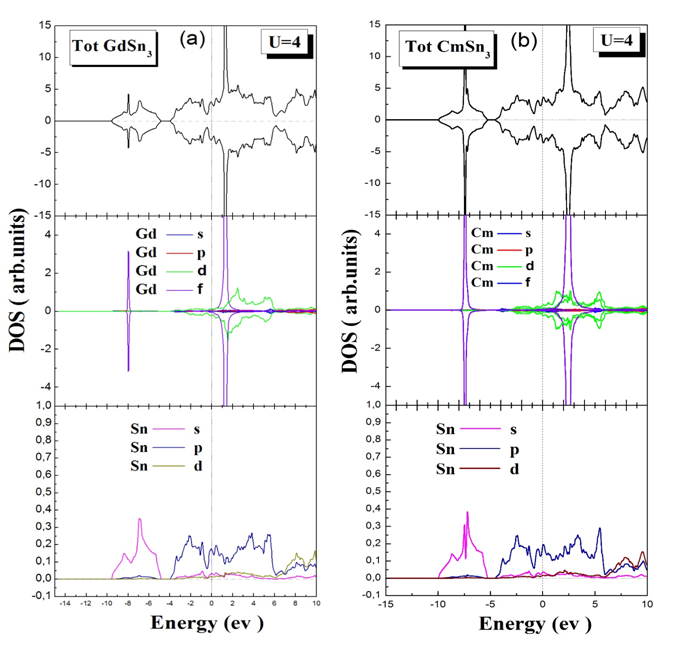}}
\caption{(Colour online) Total and partial densities of states of GdSn$_3$ and CmSn$_3$ using GGA$+U$ approach without spin-orbit coupling.}
\label{Fig:F5H}
\end{figure}

\begin{figure}[!b]
\centerline{%
\includegraphics[width=8 cm]{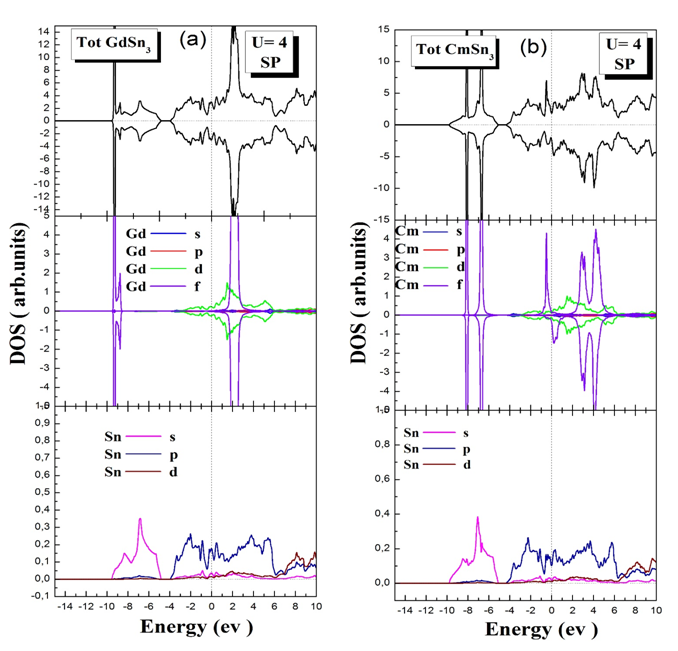}}
\caption{(Colour online) Total and partial densities of states of GdSn$_3$ using GGA$+U$ approach with spin-orbit coupling.}
\label{Fig:F6H}
\end{figure}

\begin{figure}[!b]
\centerline{%
\includegraphics[width=7 cm]{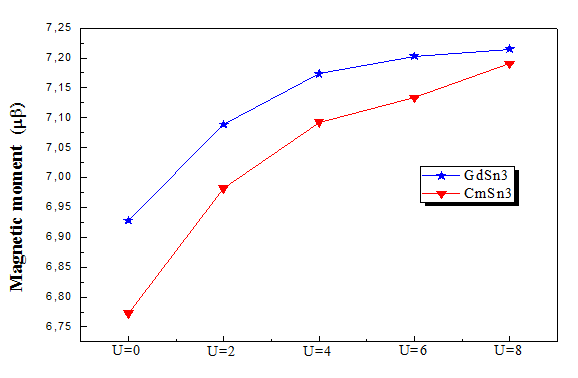}}
\caption{(Colour online) Total and partial densities of states of Gd$_{0.75}$Cm$_{0.25}$Sn$_3$, Gd$_{0.50}$Cm$_{0.50}$Sn$_3$, and Gd$_{0.25}$Cm$_{0.75}$Sn$_3$  compounds using GGA$+U$.}
\label{Fig:F7H}
\end{figure}

To have a deeper look at the electronic structure, we  displayed in figure~\ref{Fig:F5H} and in figure~\ref{Fig:F6H} the total and the partial atomic site-decomposed density of states (TDOS and PDOS) for GdSn$_3$ and CmSn$_3$, respectively, in order to analyze the contributions from the different states of Gd, Cm and Sn atoms. GGA$+U$ ($U = 4$~eV) approximation with and without the spin orbit (SO) is used to treat the exchange-correlation effects in these systems in their magnetic phases. At the beginning, we have observed that GdSn$_3$ and CmSn$_3$ exhibit a metallic behavior. 
The calculations of the total and partial density of states of GdSn$_3$ and CmSn$_3$ compounds using GGA$+U$ ($U = 4$~eV) approach without spin orbit (SO) are illustrated in figure~\ref{Fig:F5H} a and figure~\ref{Fig:F5H} b, respectively.  We should emphasize that there are two distinct structures in the density of electronic states. The first structure (from $-9.97$ to $-5.40$ eV) in the lower-lying energy side of the DOS for CmSn$_3$ consists of one important peak centered at around $-7.38$ eV, where this peak is due to Cm 5f states and with a small contribution from Sn 5s states. For GdSn$_3$, we notice a simple peak, centered at around $-8.07$ eV, in the first structure (from $-9.59$ to $-4.91$ eV), where this peak is due to Gd 4f states. The second structure [from $-4.45$ to 10 eV for CmSn$_3$ and from $-3.95$ to 10 eV for GdSn$_3$] shows a great resemblance in the  distribution of electrons in this region for both compounds, and consists of a strong peak above the Fermi level centered at around 1.34 eV and 2.37 eV for GdSn$_3$ and CmSn$_3$, respectively. This peak is due to ``f'' states. We  observed that GdSn$_3$ and CmSn$_3$ exhibit an intra-band gap in the region between $-4.91$ eV and $-3.95$ eV and between $-5.40$ eV and $-4.45$ eV, with widths of 0.96~eV and 0.95~eV, respectively. 
On the other hand, figure~\ref{Fig:F6H} a and figure~\ref{Fig:F5H} b, respectively, show the total and partial density of states of GdSn$_3$ and CmSn$_3$ compounds using GGA$+U$ ($U = 4$~eV) approach with spin orbit (SO). Considering the spin orbit effect, using the GGA$+U$ + SO approach, some changes are seen in the lower-lying energy side of the DOS. We notice the appearance of other peaks, where these peaks are due to ``f'' states of lanthanide and actinide atoms in GdSn$_3$ and CmSn$_3$, respectively. 
-
In order to estimate the effect of Cm incorporation into GdSn$_3$, we plotted in figure~\ref{Fig:F7H} the total and the partial density of states (TDOS and PDOS) for Gd$_{0.75}$Cm$_{0.25}$Sn$_3$,  Gd$_{0.50}$Cm$_{0.50}$Sn$_3$ and Gd$_{0.25}$Cm$_{0.75}$Sn$_3$, respectively, so that we can analyze the contributions of the different states of Gd, Cm and Sn atoms. GGA$+U$ ($U = 4$~eV) approximation without spin orbit (SO) is used to treat the exchange-correlation effects in these systems in their magnetic phases. We  also observed that Gd$_{0.75}$Cm$_{0.25}$Sn$_3$,  Gd$_{0.50}$Cm$_{0.50}$Sn$_3$ and Gd$_{0.25}$Cm$_{0.75}$Sn$_3$ exhibit a metallic behavior as GdSn$_3$ and CmSn$_3$, determined by the Gd-d, Sn-p, and Sn-s hybrid states that cross the Fermi level. We notice that the incorporation of Cm atoms makes the appearance of two new peaks: the first below the Fermi level centered at around $-6.25$ eV, and the second above the Fermi level centered at around 3.5 eV. However, those peaks intensities are seen to be increasing with an increase of Cm composition.  The presence of ``f'' states of Cm atoms at ~ $- 6$ eV is important and strongly localized, indicating their important contribution to the magnetic moment of Cm atoms. The different ``f'' occupancies (4f for Gd and 5f for Cm) result in different magnetic properties of these compounds. These different ``f'' occupancies are clearly observed in the valence electronic structure. The interactions between the ``f'' states [Gd (4f ) and Cm (5f)], being highly correlated and highly localized, and the ``d''   states of  Sn  have a direct effect on the previous properties. We point out that the ``d''   states contribute to the chemical bonding.

\subsubsection{ Charge densities and magnetic properties}
\begin{figure}[!t]
\centerline{%
\includegraphics[width=7 cm]{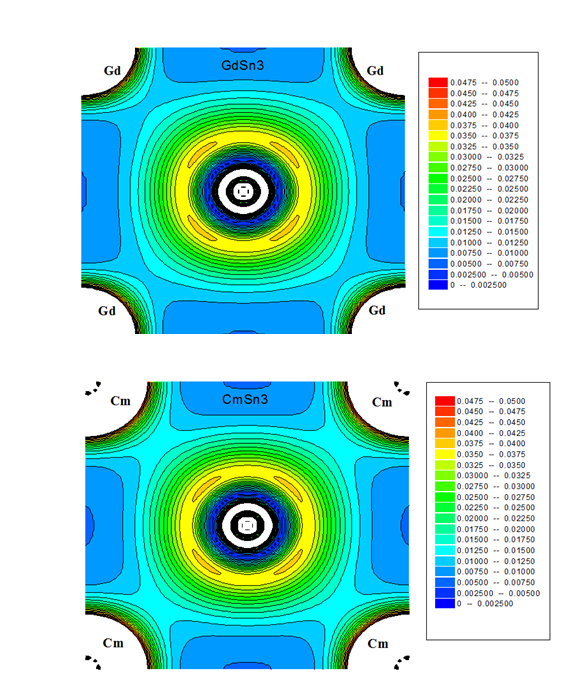}}
\caption{(Colour online) Contour plots of the valence charge distribution in the plane (110) of GdSn$_3$ and CmSn$_3$ compounds using GGA$+U$ approach.}
\label{Fig:F8H}
\end{figure}

The ionic and/or the covalent character of any compound can be related to the charge transfer between the cationic and the anionic sites; for this reason, we  calculated the total electron density of our systems GdSn$_3$ and CmSn$_3$ in the (110) plane containing Gd, Cm and Sn atoms as shown in figure~\ref{Fig:F8H}.  GGA$+U$ ($U = 4$~eV) approximation without spin orbit coupling (SO) is used to treat the exchange-correlation effects in these systems. We  observed that covalent character is predominant in the chemical bond established between Sn and Gd. The binding in CmSn$_3$ is partially covalent and ionic, since the charge density is polarized toward the Sn. These may be due to the (Pauling's) electronegativity differences between Gd and Sn atoms ($\chi$Gd$-\chi$Sn = 0.56), on the one hand, and Cm and Sn atoms ($\chi$Cm$-\chi$Sn = 0.66), on the other hand. The charge transfer corresponding to the covalent character of GdSn$_3$ is similar to that found in other researches regarding the RESn$_3$ compounds.

\begin{figure}[!t]
\centerline{%
\includegraphics[width=7 cm]{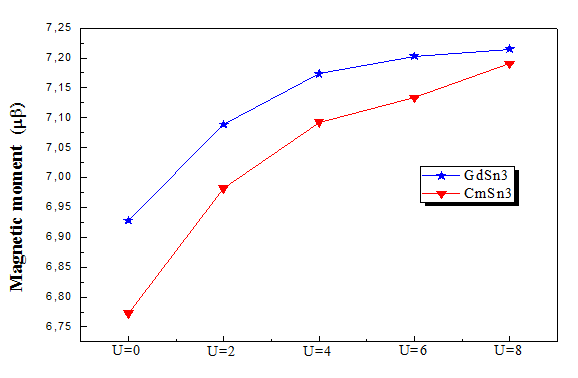}}
\caption{(Colour online) Magnetic moments per atom as a function of the effective Coulomb interaction of GdSn$_3$ and CmSn$_3$ compounds using GGA$+U$ approach.}
\label{Fig:F9H}
\end{figure}

The magnetic moment per atom of GdSn$_3$ and CmSn$_3$ compounds is shown in figure~\ref{Fig:F9H}. It is very clear from figure~\ref{Fig:F9H} that the magnetic moments increase with an increasing Hubbard potential. The calculated magnetic moments for GdSn$_3$ are underestimated compared with experimental data, but are in good agreement with the other theoretical data. The magnetic moments for Gd$_{0.75}$Cm$_{0.25}$Sn$_3$,  Gd$_{0.50}$Cm$_{0.50}$Sn$_3$ and Gd$_{0.25}$Cm$_{0.75}$Sn$_3$ compounds calculated with GGA$+U$ ($U = 4$~eV) approximation are 6.941, 6.940 and 6.940 $\mu_{\text{B}}$, respectively. We notice that the magnetic moment decreases with increasing the Cm composition.  To our knowledge, there are not available experimental or theoretical results for the magnetic moment for the studied compounds  to be compared to our theoretical results; therefore, our study can serve as a prediction for future investigations.

\section{Conclusion}
The structural, electronic and magnetic properties of the  GdSn$_3$, CmSn$_3$ and Gd$_x$Cm$_{1-x}$Sn$_3$  compounds are investigated using the FP-LAPW simulation program approach based on the density functional theory, where we have used, as an approximation for the calculation of exchange-correlation energy functional, the well-known generalized gradient approximation GGA and the GGA$+U$. From the results calculated, it is first shown that the most stable magnetic configurations of these compounds are anti-ferromagnetic type-A (AFM-A). The lattice constant for GdSn$_3$ is in reasonable agreement with the theoretical data, and is $1.47 \%$  higher than the experimental value. The lattice parameters of GdSn$_3$ and CmSn$_3$ compounds increase with  increase Hubbard potential, and their bulk modulus and its pressure derivative vary non-linearly with an increasing Hubbard potential. The band structures with and without spin orbit coupling of both compounds consistently show  a metallic nature. On the whole, the  GdSn$_3$ and CmSn compounds and their alloys show many interesting properties, including many similarities such as the presence of magnetic structures. Finally, one can point out that these compounds are of great importance and might be useful for the design of magnetic devices, frequently in high-tech applications.

\newpage
\ukrainianpart

\title%
{Першопринципні дослідження структурних, електронних i магнітних властивостей сполук XSn$_3$ (X = Gd, Cm) і Gd$_x$Cm$_{1-x}$Sn$_3$%
}
\author{M. Аднан\refaddr{label1}, Л. Джуді\refaddr{label1,label2},
       M. Марабе\refaddr{label1,label2}, M.Бушареф\refaddr{label1}, Ф. Дахман\refaddr{label2},  С. Беналья\refaddr{label1,label2}, M.~Мохтарі \refaddr{label2,label3}, Д. Рашед\refaddr{label1}}
\addresses{
\addr{label1} Лабораторія магнітних матеріалів, Університет Джіллалі Ліабе де Сіді Бель-Аббес, Алжир
\addr{label2} Інститут природничих наук і технологій, Університетський центр м. Тіссемсілт, 38000 Тіссемсілт, Алжир
\addr{label3} Оранський  інститут природничих наук і технологій ім. Мохамеда Будиафа,  31000 Оран, Алжир}
\makeukrtitle

\begin{abstract}
\tolerance=3000%
У статті досліджуються 	структурні, електронні і магнітні властивості сполук GdSn$_3$, CmSn$_3$ і Gd$_x$Cm$_{1-x}$Sn$_3$  ($x = 0.25$, $0.5$ і $0.75$) з допомогою  повно-потенціального  методу приєднаних плоских хвиль,  у рамках узагальненого градієнтного наближення$+U$. 
Властивості основного стану визначено для об'ємних сполук  GdSn$_3$, CmSn$_3$ і Gd$_x$Cm$_{1-x}$Sn$_3$, кристалізованих у структуру типу  AuCu$_3$. Обчислені структурні, електронні та магнітні властивості сполук GdSn$_3$ добре узгоджуються з iснуючими експериментальними і теоретичними даними. Встановлено, що найстійкіші магнітні конфігурації обох сполук  CmSn$_3$ і GdSn$_3$ є антиферомагнітного типу A (AFM-A) 
і проявляють металічну поведінку. Отриманий магнітний момент зменшується зі збільшенням складу  
 Cm у сполуках Gd$_x$Cm$_{1-x}$Sn$_3$. Результати показують, що сполуки  GdSn$_3$, CmSn$_3$ і Gd$_x$Cm$_{1-x}$Sn$_3$ мають деякі спільні властивості і можуть бути дуже корисними у застосуваннях.

\keywords  першопринципні дослідження, інтерметалічні сполуки, структурні властивості, електронні властивості, магнітні властивості 
\end{abstract}
\end{document}